\documentclass[11pt]{article}

\begin{document}
\topmargin 0pt
\oddsidemargin 0mm
\renewcommand{\thefootnote}{\fnsymbol{footnote}}
\begin{titlepage}

\vspace{5mm}

\begin{center}
{\Large \bf On the invariance of the speed of light} \\

\vspace{6mm} {\large Harihar Behera\footnote{E-mail: harihar@iitb.ac.in;  
 harihar@iopb.res.in } and 
  Gautam Mukhopadhyay}\footnote{E-mail: gmukh@phy.iitb.ac.in} \\
\vspace{5mm}
{\em
Department of Physics, Indian Institute of Technology, Powai, Mumbai-400076, India} \\

\vspace{3mm}
\end{center}

\vspace{5mm}
\centerline{\bf {Abstract}}
\vspace{5mm}
 The invariance of the speed of light in all inertial frames - the second
postulate of special theory of relativity (STR) - is shown to be an
inevitable consequence of the relativity principle of special theory of
relativity taken in conjunction with the homogeneity of space and time in
all inertial fames, i.e., the 1st postulate of STR. The new approach
presented here renders the learning of special theory of relativity
logically simpler, as it makes use of only one postulate.
\\

\end{titlepage}
\section{Introduction}
The two postulates of special relativity, enunciated by Einstein\cite{1}, viz.$ 
(i)$ the relativity principle: the laws by which the states of physical systems 
undergo changes are not affected, whether these changes be referred to the one 
or the other of two systems of coordinates in uniform translatory motion; and 
$ (ii) $ the postulate on the constancy of the speed of light: any ray of 
light moves in the ``stationary" system of coordinates with the determined speed 
$c$, whether the ray be emitted by a stationary or moving body; are the basis 
of a theory $-$ the special theory of relativity $-$ that has been tremendously successful in describing a wide range of phenomena, although there exist 
derivations of Lorentz transformations (LT) (that establishes a kinematical 
connection between space and time) without the postulate on the speed of 
light \cite{2,3,4,5,6,7,8,9}. However, in this communication, the invariance of the speed of 
light in all inertial frames [postulate (ii) above] is shown to be a natural 
consequence of the relativity principle [postulate (i) above] taken in 
conjuction with the homogeneity of space and time in all inertial frames. 
The approach presented here is logically simple and new.
\section{Invariance of the speed of light as consequence of the relativity principle}

Consider two inertial frames $ F $ and $ F^\prime $  which are in uniform relative 
motion along a common $X$ and $X^\prime$ axis with corresponding planes parallel 
and let the velocity of $F^\prime$ with respect to $F$ be ${\bf v}$ along the 
positive $X$-axis of $F$. An event may be characterized by specifying the coordinates 
 $ (x, y, z, t) $ of the event in $ F $ and the same event is characterized by the coordinates $ (x^\prime, y^\prime, z^\prime, t^\prime) $ in $ F^\prime $. Let us proceed to find a transformation between $ ( x, y, z, t) $ and $ ( x^\prime, y^\prime, z^\prime, t^\prime) $ and then deduce from it Einstein's postulate on the velocity of light in special relativity. Suppose that at the instant the origins $ 0 $ and $ 0^\prime$ coincide, we let the clocks there  to read $ t = 0 $ and $ t^\prime = 0 $ respectively. The homogeneity of space and time  in inertial frames requires that the transformations must be linear (see for example \cite{10}), so that the simplest form they can take is\\                        
 \begin{equation} x' = k (x - vt);\>\>\>   y' = y;\>\>\>   z' = z;\>\>\>  t' = lx + mt                                                                        
 \end{equation} \\
In order to determine the values of the three co-efficients $ k,l,$ and $ m $, 
let us imagine that at instant when the origins of $F$ and $F^\prime$ coincide 
(i.e., at $t = t^\prime = 0$), a light wave left a point source kept stationary at 
the origin of $F$. Since the space is homogeneous and isotropic in $F$, the light 
originating from the stationary point source in $F$ would spread uniformly in all directions with the determined speed $c$ (say). As the source is moving with velocity 
($-{\bf v}$) with respect to $F^\prime$, let us assume that the speed of light in 
$F^\prime$ as $c^\prime$, assuming we are unaware of Einstein's postulate on the 
constancy of the speed of light in different inertial frames. Some time after the 
moment $t = t^\prime = 0$, when the clocks of $F$ read a time $t$, suppose that 
the clocks of $F^\prime$ read a time $t^\prime$.  The wavefront of the light wave 
looks spherical in $F$, because the wave originating from a stationary point source 
moves in the homogeneous and isotropic space of $F$. But we are yet uncertain about 
the shape of the same wavefront in $F^\prime$, because we know not yet whether 
$c^\prime$ is uniform in all directions of $F^\prime$ or not. Now let us consider 
the space and time coordinates of an arbitrary point on the same wavefront observed in $F$ 
as $(x, y, z, t)$ and the space and time coordinates of the same point on the same 
wavefront observed in $F^\prime$ as $(x^\prime, y^\prime, z^\prime, t^\prime)$. So 
the space coordinates $(x, y, z)$ of the wavefront in $F$ differ from point to 
point of the wavefront and similarly the space coordinates $(x^\prime, y^\prime, 
z^\prime, t^\prime)$ of the wavefront in $F^\prime$ also differ from point to point. 
But the time coordinate $t$ is the same for all points of the wavefront in $F$ and 
similarly the time coordinate $t^\prime$ is the same for all points of the wavefront 
in $F^\prime$ \cite{1}. Then for an arbitrarily chosen point on the wavefront 
characterized by the space-time coordinates $(x, y, z, t)$ and $(x^\prime, y^\prime, 
z^\prime, t^\prime)$ respectively, we must have the following realtions for the 
same point on the same wavefront as observed in $F$ and $F^\prime$: 

\begin{equation} 
x^{2} + y^{2} + z^{2} = c^{2}t^{2}                            
 \end{equation}                                                                 
 \begin{equation}
 {x^\prime}^{2} +{y^\prime}^{2} +{z^\prime}^{2} ={c^\prime}^{2}{t^\prime}^ {2}                         
 \end{equation}                                                                
 Substituting the transformations (1)in (3), one gets: 
\begin{equation} 
(k^{2} - l^{2}{c^\prime}^{2})x^{2} + y^{2} + z^{2} -2(lm{c^\prime}^{2} + k^{2}v)xt =
(m^{2}{c^\prime}^{2} - k^{2}v^{2})t^{2}  
\end{equation}                                           
For Eq.(4) to agree with Eq.(2), we must have
\begin{eqnarray} 
(i) && k^{2} - l^{2}{c^\prime}^{2} = 1, \>\>\>\> {\mbox {or}} \>\>\>\> l^2 = (k^2 - 1)/{c^\prime }^2 \nonumber \\
(ii) && m^{2}{c^\prime}^{2} - k^{2}v^{2} = c^{2}, \>\>\>\> {\mbox {or}} \>\>\>\>  m^2 = (c^2 + k^2v^2)/{c^\prime}^2  \nonumber \\ 
(iii) && lm{c^\prime}^{2} + k^{2}v = 0,  \>\>\>\> {\mbox {or}} \>\>\>\> lm = -k^2v/{c^\prime}^2  
\end{eqnarray}       
\noindent
Equations (5) when solved for $ k,l $ and $ m $ yield
\begin{eqnarray} 
(i) && k = (1 - {v^{2}}/{c^{2}})^{-1/2} \nonumber \\
(ii) && l = -(v/cc')k \nonumber \\
(iii) && m = (c/c')k
\end{eqnarray}  
With these values of $k, l$ and $ m $, the transformations
(1) can be represented in the following matrix form, viz.,     
\begin{equation}\left( \begin{array}{c} x'\\ y'\\ z'\\ t'
    \end{array}\right) = A \left( \begin{array}{c} x \\ y \\ z \\ t
    \end{array}\right) ,\>\>\>\> {\rm where} \   A = \left(
    \begin{array}{cccc} k & 0 & 0 & -kv \\ 0 & 1 & 0 & 0 \\ 0 & 0 & 1
      & 0 \\ -kv/cc' & 0 & 0 & kc/c' \end{array} \right)      
\end{equation}

The inverse transformations for $ x, y, z, t $, are then given by
\begin{equation}\left( \begin{array}{c} x \\ y \\ z \\ t
\end{array}\right) = A^{-1} \left( \begin{array}{c} x' \\  y' \\ z'\\
  t' \end{array}\right)  ,\>\>\>\> {\rm where} \  A^{-1} = \left(
\begin{array}{cccc} k  &  0  & 0&  kvc'/c  \\  0  &  1  &  0  &  0  \\
  0  &  0  &  1 & 0  \\  kv/c^{2}  &  0  &  0 & kc'/c  \end{array}
\right)  
\end{equation} 
Hence the transformation for $ x $ becomes
\begin{equation} x = k (x' + vc't'/c)
\end{equation}
But relativity principle demands that the transformation for $ x $ must be given 
by \cite{10}:
\begin{equation} x = k(x' + vt')
\end{equation}
Hence for Eq.(9) to be in accord with the relativity principle (so with Eq.(10)),
we must have $c=c^\prime$. 
\section{Conclusion}
We have, thus, shown that the constancy of the speed of light in all inertial frames is an inevitable consequence of the realtivity principle taken in conjunction with the 
homogeneity of space and time in all inertial frames. The present approach renders 
the theory of special relativity logically simpler as it makes use of only one postulate. The idea of constructing a relativity theory by using only the relativity 
principle has also been discussed by Ritz, Tolman and Pauli \cite{11}, but the 
present approach here is new and simpler and has been given as preliminary 
report in \cite{12}.\\


\end{document}